\documentclass[twocolumn]{article}
\usepackage{amsmath, amsfonts}
\usepackage{float}
\usepackage{booktabs, siunitx, caption}
\usepackage{hyperref}
\usepackage{xcolor}
\usepackage{graphicx}
\usepackage[linesnumbered, ruled, vlined]{algorithm2e}

\title{\LARGE \bfseries Hacking, The Lazy Way: LLM Augmented Pentesting}

\author{
\begin{minipage}[t]{0.45\textwidth}
\centering
Dhruva Goyal \\
Chief Executive Officer, BugBase Security \\
\texttt{dhruva@bugbase.ai}
\end{minipage}\hfill
\begin{minipage}[t]{0.45\textwidth}
\centering
Aditya Peela \\
Chief Technology Officer, BugBase Security \\
\texttt{aditya@bugbase.ai} \\
Department of ICT, MIT, Manipal University
\end{minipage} \\[1.5em]
\begin{minipage}[t]{0.45\textwidth}
\centering
\BlankLine
Sitaraman Subramanian \\
Chief Information Officer, BugBase Security \\
\texttt{sitaraman@bugbase.ai} \\
Department of ICT, MIT, Manipal University
\end{minipage}\hfill
\begin{minipage}[t]{0.45\textwidth}
\centering
\BlankLine
Nisha P Shetty \\
\texttt{nisha.pshetty@manipal.edu} \\
Department of ICT, MIT, Manipal University
\end{minipage}
}
\date{}

\begin{document}

\twocolumn
\maketitle

\begin{abstract}
In our research, we introduce a new concept called ``LLM Augmented Pentesting'' demonstrated with a tool named ``Pentest Copilot,'' that revolutionizes the field of ethical hacking by integrating Large Language Models (LLMs) into penetration testing workflows, leveraging the advanced GPT-4-turbo model. Our approach focuses on overcoming the traditional resistance to automation in penetration testing by employing LLMs to automate specific sub-tasks while ensuring a comprehensive understanding of the overall testing process.

Pentest Copilot showcases remarkable proficiency in tasks such as utilizing testing tools, interpreting outputs, and suggesting follow-up actions, efficiently bridging the gap between automated systems and human expertise. By integrating a ``chain of thought'' mechanism, Pentest Copilot optimizes token usage and enhances decision-making processes, leading to more accurate and context-aware outputs. Additionally, our implementation of Retrieval-Augmented Generation (RAG) minimizes hallucinations and ensures the tool remains aligned with the latest cybersecurity techniques and knowledge. We also highlight a unique infrastructure system that supports in-browser penetration testing, providing a robust platform for cybersecurity professionals. Our findings demonstrate that LLM Augmented Pentesting can not only significantly enhance task completion rates in penetration testing but also effectively addresses real-world challenges, marking a substantial advancement in the cybersecurity domain.
\end{abstract}

\vspace{0.5em} 

\section{Introduction}
The rapid advancement of Transformer based deep neural networks like Large Language Models (LLMs) have significantly transformed numerous industries by enabling machines to understand, generate, and interact with human language in unprecedented ways.\cite{tann2023usinglargelanguagemodels} From automating customer service to assisting in code generation in complex codebases, LLMs have demonstrated their potential as general-purpose reasoning engines capable of augmenting complex human workflows. Their ability to contextualize, summarize, and synthesize vast volumes of data has made them increasingly relevant in domains requiring both breadth and depth of knowledge.

In the domain of cybersecurity, penetration testing (pentesting) serves as a cornerstone of proactive defense, simulating attacks to identify vulnerabilities before malicious actors can exploit them. However, traditional pentesting is heavily reliant on the expertise and intuition of skilled professionals. This human dependency has created a bifurcated market: high-quality, manual pentests that are expensive and slow, and low-cost, automated tools that are often mostly compliance-driven and fail to uncover deeper vulnerabilities. Consequently, many organizations find themselves performing pentests as a checkbox exercise, gaining little real security value and remaining exposed to data breaches and regulatory penalties.\cite{b18}

As the complexity of digital systems grows, the demand for scalable, intelligent, and cost-effective security solutions has intensified. While artificial intelligence has been improving various aspects of cybersecurity, such as threat detection or anomaly analysis, its role in penetration testing remains nascent and underexplored. Most existing automation tools in pentesting are limited to scripted tasks and lack the adaptability required for dynamic and context-aware exploration of systems.

This paper introduces LLM-Augmented Pentesting \cite{tann2023usinglargelanguagemodels,Happe_2023}, a novel approach aimed at bridging this gap. By integrating the reasoning capabilities of LLMs with the practical knowledge and tools of pentesters, it proposes a new paradigm for intelligent pentesting assistance. To demonstrate this approach, we developed a tool named 'Pentest Copilot,' which integrates LLMs to assist with various stages of the penetration testing process

Human security testing workflows are enhanced by Pentest Copilot through the automation of routine tasks such as documentation lookup, tool orchestration, and the generation of exploit strategies. Unlike traditional automation tools, the expansive, generalized knowledge of LLMs is leveraged by Pentest Copilot to offer contextual suggestions, maintain awareness of the testing process, and provide insights that are aligned with the nuanced and evolving nature of penetration testing.

Through this hybrid approach, access to high-quality pentesting is aimed to be democratized---made more efficient, effective, and affordable---without compromising the expertise on which human-driven assessments rely. The standards of assurance in a rapidly evolving threat landscape are being redefined by the fusion of LLMs with cybersecurity practices \cite{peng2023impactaideveloperproductivity}.

\section{Research Survey}
Recent developments in LLMs have inspired a wave of research exploring their applicability to cybersecurity tasks. The surge of interest has resulted in diverse approaches ranging from educational benchmarking and static code analysis to modular frameworks and automated attack agents. However, most existing works, while valuable in their own right, fall short of integrating LLMs into live, dynamic, and context-rich penetration testing workflows. This section provides a structured overview of the existing body of work, categorized by thematic focus, and highlights the critical gaps that remain unaddressed in the space of real-world pentesting automation.

\subsection{Educational and Static Evaluation of LLMs}
Several early efforts focused on evaluating LLM capabilities in controlled educational contexts or static environments, such as Capture-The-Flag (CTF) challenges and cybersecurity certification questions.

Tann et al. (2023)\cite{tann2023usinglargelanguagemodels} investigated how GPT-based models performed in solving CTF problems and answering knowledge-based security certification questions. Their study highlighted the ability of LLMs to excel in factual question answering and certain challenge categories. However, the study was confined to academic evaluation settings and did not explore how LLMs could be integrated into real-world pentesting workflows. Notably, it did not address the dynamic execution of commands, orchestration of tools, or the challenges posed by unstructured inputs such as binaries or configuration files---tasks that are commonplace in operational pentesting environments.

Pratama et al. (2024) \cite{Pratama_2024} introduced CIPHER, a domain-specific LLM fine-tuned on over 300 expert pentesting write-ups. The paper proposed a novel FARR Flow benchmark (Findings, Actions, Reasoning, Results) to evaluate LLMs in pentesting-like scenarios. While CIPHER made meaningful strides in guiding novices with structured and explainable recommendations, the interaction style remained chat-assistant-oriented and largely static. It did not delve into tool orchestration, real-time shell interactions, or the chaining of live tools across complex environments.

Zhang et al. (2023)\cite{zhang2023doesllmgeneratesecurity} explored prompt engineering strategies to guide ChatGPT-4 in generating JUnit-based security tests that target applications with vulnerable third-party dependencies. Although the model outperformed existing tools in exploit generation and test validity, the setup was highly curated and offline. It required expert-provided vulnerability metadata and focused solely on static, Java-based applications---without addressing generalized, live, or dynamic pentesting workflows.

These studies collectively showcase the knowledge-processing capabilities of LLMs but largely within simulated or educational environments. While they provide strong evidence of the model's reasoning and understanding, they do not address the operational complexity and live-tool interaction required in real-world penetration testing.

\subsection{Modular LLM Frameworks with Limited Realism or Human Dependence}
A more recent line of research has focused on developing modular frameworks that use LLMs to automate parts of the pentesting workflow. These systems attempt to mirror real-world testing dynamics but often fall short of full autonomy, extensibility, or interaction with production-grade systems.

Deng et al. (2024)\cite{deng2024pentestgptllmempoweredautomaticpenetration} presented PentestGPT, a modular LLM-based framework designed to simulate penetration testing team dynamics. Using a Pentesting Task Tree (PTT), PentestGPT maintains task context across multiple steps and integrates tool usage into a semi-automated workflow. Their evaluation on benchmarks from HackTheBox and VulnHub showcased the framework's ability to decompose complex tasks and reason through multi-step engagements. However, execution remains human-in-the-loop, with no direct system integration, shell orchestration, or live tool chaining. The framework also does not support dynamic parsing of binaries or configuration files---key requirements for generalized real-world pentests.

AlShehri et al. (2024) \cite{alshehri2024breachseekmultiagentautomatedpenetration} introduced BreachSeek, a multi-agent pentesting automation platform powered by LangChain and LangGraph. The platform assigns specialized LLM agents (e.g., Supervisor, Pentester, Evaluator) to autonomously discover vulnerabilities and execute attacks in Metasploitable-like environments. While the architectural modularity is promising, the system operates in localized, homogenous lab setups and lacks integration with real-time toolchains, arbitrary file formats, or enterprise-grade environments. Moreover, BreachSeek does not offer plugin extensibility or perform dynamic binary/config parsing---both essential for scalable automation.

Xu et al. (2024) \cite{xu2024autoattackerlargelanguagemodel} developed AUTOATTACKER, an LLM-guided system that automates hands-on-keyboard attack chains post-breach. Their pipeline includes a summarizer, planner, navigator, and experience manager, with GPT-4 at the core and RAG for context enhancement. AUTOATTACKER is notable for completing multi-step attacks across 14 real-world tasks without human intervention. However, the scope is limited to pre-configured virtual labs with fixed objectives. It does not generalize to new targets, lacks a plugin-based extensibility model, and avoids direct shell command orchestration across diverse infrastructures.

These frameworks demonstrate ambitious attempts at structuring and automating pentesting workflows, but are often constrained by closed testbeds, static targets, lack of real-time plugin chaining, and absence of generalized tooling interoperability across varying systems.

\subsection{Task-Specific or Narrowly Focused LLM Applications}
Several papers focus on specific sub-tasks within the broader security domain---such as code patching, exploit generation, or unit test synthesis---but are not designed to support full-stack black box penetration testing workflows.

Pearce et al. (2022) \cite{pearce2022examiningzeroshotvulnerabilityrepair}  explored LLMs like Codex and Jurassic-1 for zero-shot vulnerability repair, evaluating their performance on real-world and synthetic security bugs. Their research provided valuable insights into prompt engineering, model parameter tuning, and dataset design for code patches. However, their setup is tailored to isolated code-level tasks, without any connection to external systems, dynamic file inputs, or runtime environments---thus lacking any contribution to live pentesting automation.

Zhang et al. (2023)\cite{zhang2023doesllmgeneratesecurity} similarly focused on Java applications and JUnit-based exploit test generation, further illustrating the narrow task scope of many LLM cybersecurity applications. These studies are useful for isolated workflows (e.g., SAST), but are not extensible to operational, system-level pentesting.

These works underscore the potential of LLMs in focused security domains, but reveal a clear gap in their applicability to holistic, real-world penetration testing scenarios.

\subsection{Pre-LLM RL-Centric Automation Efforts}
Before the mainstream adoption of LLMs, researchers explored other AI paradigms to automate parts of penetration testing.

Schwartz (2018) \cite{schwartz2019autonomouspenetrationtestingusing} proposed using reinforcement learning (RL) to simulate penetration testing as a Markov Decision Process (MDP), training agents via tabular and neural Q-learning inside abstracted networks. The framework included a lightweight Network Attack Simulator (NAS) to enable experimentation. While conceptually innovative, the system remained academic, operating in synthetic environments with abstracted vulnerabilities. It lacked real-world tooling like Metasploit or Burp Suite, had no language-based reasoning capabilities, and did not account for parsing or interacting with real system outputs (e.g., logs, configs, binaries, or live shells).

This line of work is important from a historical and theoretical standpoint but falls short of the contextual, reasoning-driven, tool-integrated workflows enabled by LLMs today.

\subsection{Recurring Limitations in Existing Work}
Across these diverse approaches, several critical limitations emerge repeatedly:

\begin{itemize}
\item Static or Simulated Environments: Most systems are tested on known environments (CTFs, VulnHub, HackTheBox) and are not built to adapt to arbitrary, unknown systems.
\item Lack of Real-Time Tool Chaining: Few, if any, explore dynamic chaining of tools or automated follow-up commands based on live output.
\item No Plugin-Based Extensibility: Current frameworks lack modular plugin architectures that allow seamless execution of common pentesting tools.
\item Human-in-the-Loop Dependency: Many systems rely on human operators to validate or execute commands, limiting the automation potential.
\item Limited Input Handling: Parsing and reasoning over binaries, configuration files, media, or shell output is either absent or done manually.
\item No Live Shell or Command Orchestration: Most tools do not facilitate live sessions, shell access, or session-state awareness across a full engagement.
\end{itemize}

These collective shortcomings point to the absence of a truly operational-grade, LLM-driven pentesting assistant that can operate in real-time, adapt dynamically, orchestrate tools across varied targets, and handle diverse input formats.

To guide the investigation into LLM-Augmented Pentesting and the development of Pentest Copilot, following research questions are posed:

\begin{itemize}
\item RQ1: How does the performance of different GPT-based LLMs (GPT-3.5, GPT-4, GPT-4-Turbo, GPT-4o) vary in real-world pentesting scenarios?
\item RQ2: How can Retrieval-Augmented Generation (RAG) be used to reduce hallucinations and maintain up-to-date knowledge of pentesting tools and techniques?
\item RQ3: How can non-textual inputs commonly encountered during penetration testing, such as binaries, configuration files, or media, be effectively transformed into formats that LLMs can process and reason over?
\item RQ4: What infrastructure and system design are necessary to support secure, scalable, and interactive LLM-assisted pentesting?
\item RQ5: To what extent does Pentest Copilot improve task completion time, workflow efficiency, and output quality for penetration testers?
\end{itemize}

\section{Methodology}

\subsection{Large Language Model}

\subsubsection{Selecting a class of models}
The research indicated that among various LLMs available, the GPT lineage of models from OpenAI stood out for its extensive knowledge on security toolings and use cases \cite{deng2024pentestgptllmempoweredautomaticpenetration,Happe_2023}. Its global knowledge, large training dataset, encompassing a wide array of cybersecurity papers and topics.

The goal was to leverage a well-trained model capable of addressing cybersecurity and pentesting queries through guided instructions, before proceeding to finetune an in-house open source model.

\subsubsection{Surface Evaluation of GPT Models}
Evaluation on the GPT lineage models was performed, starting out with GPT-3.5-Turbo, GPT-4 and GPT-4-turbo \cite{b7,b8}.

\begin{itemize}
\item \textbf{GPT-3.5-Turbo}: GPT-3.5-Turbo is notably fast in generating responses, making it appealing for quick pentesting tasks. However, it often struggles to deliver accurate results in more complex scenarios. The model has difficulty keeping track of details over long interactions, which is a significant limitation in penetration testing where past findings matter. Its reliance on older data also means it may not always reflect the latest tools or techniques, making it less ideal for detailed or advanced pentesting needs.

\item \textbf{GPT-4}: GPT-4 offers improved accuracy compared to earlier models, performing well in complex pentesting situations where understanding the context of previous steps is key. It provides detailed and relevant information, which is valuable for thorough testing. However, its responses are not always structured in the expected format, which can hinder its effectiveness. Additionally, GPT-4 is much slower than other models, reducing its practicality for tasks requiring quick turnaround in dynamic pentesting environments.

\item \textbf{GPT-4-Turbo}: GPT-4-Turbo stands out as the strongest choice for penetration testing among the models evaluated. It handles complex tasks with high accuracy, maintaining context effectively even during extended testing sessions. Its speed is reasonable, striking a good balance between responsiveness and reliability. Benefiting from more recent data, GPT-4-Turbo aligns well with current pentesting practices, making it suitable for demanding and intricate pentesting projects.

\item \textbf{GPT-4o}: GPT-4o delivers a balanced performance, offering decent accuracy for penetration testing tasks. It manages context well, allowing it to reference earlier findings during tests, which is helpful for moderately complex scenarios. With access to up-to-date information, it stays relevant for modern pentesting needs. However, its response speed is slower than some lighter models, which may be a drawback when rapid feedback is essential. GPT-4o is a solid option for testers seeking reliability without the highest level of precision.

\item \textbf{GPT-4o-mini}: GPT-4o-mini is the fastest model evaluated, making it excellent for pentesting situations that demand quick responses. Despite its speed, it sacrifices some accuracy, struggling with more intricate tasks compared to larger models. It can still use recent techniques but may lose track of details in longer testing sessions. Its compatibility with pentesting tools is strong, matching more advanced models in this regard. GPT-4o-mini is best suited for straightforward pentesting tasks where speed is prioritized over deep accuracy.
\end{itemize}

\subsubsection{Testbenching Framework and Evaluation}
A testbenching framework was developed to mimic real-world pentesting scenarios, spanning vulnerability assessments through post-exploitation. A purposefully vulnerable “boot2root” server \cite{b11,b12,b13} was hosted to evaluate each model’s capability to achieve Remote Code Execution. From initial reconnaissance to code execution, thirty manually curated test cases—drawn from bug-bounty reconnaissance, XSS, SQL injection, privilege escalation, and other common workflows—reflect the complexity of real engagements.

Each test case was defined to include:
\begin{itemize}
  \item A structured JSON prompt containing instructions, goals, and constraints.
  \item Available plugins (see “Dissecting Prompts” for full definitions), e.g.\ \texttt{run\_bash}, \texttt{msfvenom\_payload}, \texttt{netcat\_listener}, \texttt{google}.
  \item Expected output fields: \texttt{expected\_plugins}, \texttt{expected\_content}, and \texttt{binaries\_used}.
\end{itemize}

Models were evaluated on four quantitative metrics:
\begin{itemize}
  \item \textbf{Structural Accuracy (\%)}: Conformance to the predefined JSON schema, ensuring outputs remain parseable.
  \item \textbf{Functional Correctness (\%)}: Calculated as TP/(TP+FP), where TP represents correctly identified techniques, True Positives, and FP represents incorrect identifications, False Positives.
  \item \textbf{Command Accuracy (\%)}: Proportion of correctly suggested commands.
  \item \textbf{Plugin Validity (\%)}: Correct selection and syntax of expected plugins.
\end{itemize}
Additionally, the following were measured:
\begin{itemize}
  \item \textbf{Consistency (\%)}: Stability of outputs across repeated runs.
  \item \textbf{Average Response Time (s)}: Average time to get output.
\end{itemize}

The testbenching and evaluation framework is available as open-source at: \url{https://github.com/Hackerbone/HackerLLMBench/}

Table \ref{tab:performance1} presents structural and functional results;  
Table \ref{tab:performance2} reports consistency and command accuracy;  
Table \ref{tab:performance3} provides plugin validity and average response times.

\newcolumntype{L}{>{\raggedright\arraybackslash}X} 
\newcolumntype{C}{>{\centering\arraybackslash}X}   
\sisetup{detect-all,table-format=3.2}

\newcolumntype{P}{>{\raggedright\arraybackslash}p{0.20\linewidth}} 
\newcolumntype{Y}{>{\centering\arraybackslash}X}                 
\sisetup{detect-all,table-format=3.2}
\setlength{\tabcolsep}{6pt} 

\begin{table}[H]
  \centering
  \caption{Structural and Functional Performance}
  \begin{tabular}{|p{2.6cm}|p{2cm}|p{2cm}|}
    \hline
    \textbf{Model}       & \textbf{Structural Accuracy (\%)} & \textbf{Functional Correctness (\%)} \\
    \hline
    GPT-3.5-Turbo        & 100 &  40 \\
    GPT-4                &   0 &   0 \\
    GPT-4-Turbo          & 100 &  60 \\
    GPT-4o               & 100 &  50 \\
    GPT-4o-mini          & 100 &  40 \\
    \hline
  \end{tabular}
  \label{tab:performance1}
\end{table}

\begin{table}[H]
  \centering
  \caption{Consistency and Command Accuracy}
  \begin{tabular}{|p{2.6cm}|p{2cm}|p{2cm}|}
    \hline
    \textbf{Model}       & \textbf{Consistency (\%)} & \textbf{Command Accuracy (\%)} \\
    \hline
    GPT-3.5-Turbo        &  40 &  40 \\
    GPT-4                &   0 &   0 \\
    GPT-4-Turbo          &  60 &  60 \\
    GPT-4o               &  50 &  50 \\
    GPT-4o-mini          &  40 &  40 \\
    \hline
  \end{tabular}
  \label{tab:performance2}
\end{table}

\begin{table}[H]
  \centering
  \caption{Plugin Validity and Average Response Time}
  \begin{tabular}{|p{2.6cm}|p{2cm}|p{2cm}|}
    \hline
    \textbf{Model}       & \textbf{Plugin Validity (\%)} & \textbf{Avg. Response Time (s)} \\
    \hline
    GPT-3.5-Turbo        &  50 &  3.47 \\
    GPT-4                &   0 & 20.99 \\
    GPT-4-Turbo          &  80 &  7.11 \\
    GPT-4o               &  80 &  4.52 \\
    GPT-4o-mini          &  80 &  3.41 \\
    \hline
  \end{tabular}
  \label{tab:performance3}
\end{table}

\subsection{Prompt Engineering}
Prompt engineering plays an important role in LLM Augmented Pentesting, ensuring that the model understands and responds to cybersecurity-related queries with precision. By crafting specific, context-aware prompts, the model can interpret technical language and ethical hacking related terminologies effectively \cite{deng2024pentestgptllmempoweredautomaticpenetration,Happe_2023}.

\subsubsection{Jailbreaking GPT}
By design, GPT is not configured to specifically cater to pentesting queries. Its general-purpose design meant that it provided broad information across various domains, but due to security reasons and constraints placed by OpenAI \cite{b7,b8}, the model is aligned to not respond to a pentesting query.

The key was to instruct GPT to assume the role of a ``Penetration testing assistant, collaborating with a security researcher'' \cite{liu2024jailbreakingchatgptpromptengineering}. The instruction was given to GPT in its system prompt which is a constraint that it will follow at all times while returning a response. The different types of prompts are elaborated in the ``Dissecting Prompts'' section.

This shifted the model's operational framework from a generalist to a specialist, focusing its vast knowledge base specifically towards pentesting.

\subsubsection{Dissecting Prompts}
We have utilized the Chat Completions API of the GPT4-turbo model. The primary input for Chat Completions API is the ``messages'' parameter, which is an array of message objects. Each message object has a ``role'' (either ``system,'' ``user,'' or ``assistant'').

\paragraph{System/Instruction Prompt}
The system prompt at the initiation of Pentest Copilot explicitly outlines the objectives, enumerates constraints, and provides context regarding the configured tools for utilization in commands. This is crucial to prevent inadvertent tool name hallucinations and inaccurate information by the LLM.

\paragraph{User Prompt}
In the user prompt, we dynamically adjust the instructions based on whether the user provides previous testing information for the pentest target or not. If the user provides previous testing details, we skip adding the prompt for the preliminary testing phase. However, if no previous testing information about the target is provided, the prompt with either a Domain or IP scan is added. The prompt also outlines the expected JSON response format for output and sets constraints, such as recommending a single command if possible. Emphasis is placed on prioritizing user input to ensure the pentest aligns with their preferences, avoiding autonomous decisions by the model. Additionally, if no target information is provided, Pentest Copilot asks for more details about the target.

For tasks such as executing commands and creating payloads, a set of plugins are directly specified in the prompt. Here's an overview of some of these plugins:

\begin{itemize}
\item \textbf{Web Search Plugin}: Enables the LLM to conduct web searches for discovering new exploits or methodologies.
\item \textbf{Run Bash Plugin}: Allows the LLM to execute bash commands including open source pentest tools for various tasks.
\item \textbf{Generic Response Plugin}: This plugin gives a text response that contains insights from the LLM when specific tasks cannot be completed using plugins. Moreover, Pentest Copilot can execute this plugin to request additional information from the user.
\item \textbf{Netcat Listener Plugin}: Initiates a Netcat listener on a specified port, creating a listener for a reverse shell connection.
\item \textbf{Generate Payload Plugin}: Generates an exploit payload using the metasploit framework \cite{b9}.
\end{itemize}

\begin{figure}[H]
\centering
\includegraphics[width=0.9\columnwidth]{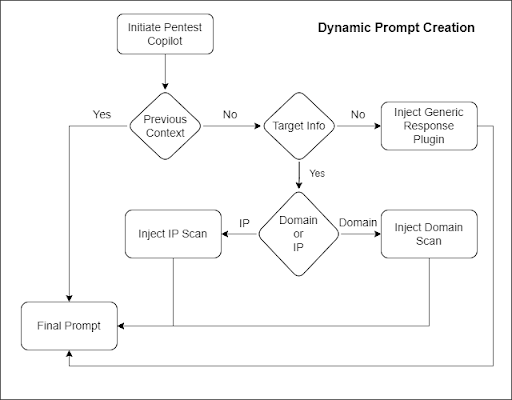}
\caption{Flowchart on Dynamic Prompt Creation where new target information and previous context is injected into the user prompt based on the current state of pentest}
\label{fig1}
\end{figure}

These plugin specifications are incorporated directly into the prompt, providing clear instructions to the model on when and how to employ each plugin.

\subsubsection{Prompt Components}
Pentest Copilot has three sequential prompts per pentest loop. A pentest loop is a single iterative cycle that includes command generation, summarization, and to-do list updates to systematically advance the penetration testing process. It includes:
\begin{enumerate}
    \item \textbf{Command Generation Prompt}
    \item \textbf{Summarization Prompt}
    \item \textbf{To-Do Update Prompt}
\end{enumerate}

Each prompt bundles only the components required for its task:

\subsection*{1. Command Generation Prompt (approximately 1,350 tokens)}
\begin{itemize}
    \item \textbf{Core System Instructions (250 tokens)}: Defines the ethical hacking assistant persona, establishing guidelines for responsible and secure behavior.
    \item \textbf{Plugin Descriptions (600 tokens)}: Details available plugins, such as \texttt{run\_bash} or \texttt{google}, enabling specific actions within the system.
    \item \textbf{Response Format Guidelines (500 tokens)}: Specifies the JSON structure for responses, including field constraints to ensure consistency and clarity.
\end{itemize}

\subsection*{2. Summarization Prompt (approximately 1,800 tokens)}
\begin{itemize}
    \item \textbf{JSON Schema Constraints (500 tokens)}: Enforces a standardized output format to maintain uniformity across summaries.
    \item \textbf{Command Generation Output (500 tokens)}: Includes the latest command executed and its unprocessed output for reference.
    \item \textbf{Summary Template (800 tokens)}: Provides a structured format for summarizing outcomes, capturing essential state changes and key results.
\end{itemize}

\subsection*{3. To-Do Update Prompt (approximately 1,900 tokens)}
\begin{itemize}
    \item \textbf{Previous To-Do Checklist (800 tokens)}: Carries forward outstanding objectives from prior iterations to ensure continuity.
    \item \textbf{History of Recent Actions (600 tokens)}: Maintains context by summarizing recent activities and their outcomes.
    \item \textbf{Update Instructions}: Guides the process of revising the To-Do checklist based on the latest summary, aligning with current objectives.
    \item \textbf{Response Structure JSON Guidelines (500 tokens)}: Defines the JSON format for parsing To-Do updates, ensuring structured and consistent output.
\end{itemize}

By cycling through these three focused prompts—command generation, result summarization, and task-list update—only the minimal necessary context is carried forward at each step. This targeted allocation preserves token budget, maintains clarity, and enables deep, multi-step reasoning across reconnaissance, enumeration, exploitation, and post-exploitation phases (Fig.~\ref{fig2}).

\subsection{Step Chaining Approach for Token Size Optimization}
Managing the 4,096-token context limit imposed by the OpenAI Chat Completions API for inference was a key challenge in developing Pentest Copilot. To ensure detailed, step-wise guidance without exceeding this limit, a step-chaining strategy—akin to the chain-of-thought approach \cite{wei2023chainofthoughtpromptingelicitsreasoning}—was implemented. Complex pentesting workflows are split into discrete phases (e.g.\ reconnaissance, enumeration, exploitation), with each phase’s output distilled and carried forward as compressed context for the next API call. This method preserves essential information while staying within token bounds and enables deep, multi-step reasoning across the entire engagement.

The step-chaining approach addresses key limitations of using a single large prompt in the pentesting workflow:
\begin{itemize}
    \item \textbf{Hallucinations}: The model may generate incorrect or fabricated information during a phase, which, if carried forward, can compromise the accuracy of subsequent steps in the pentesting workflow.
    \item \textbf{Context Drops}: The model may ignore or miss critical information from previous phases, leading to incomplete context and potential errors in multi-step reasoning.
    \item \textbf{Token Size Issues}: Compressing context to fit within the 4,096-token limit may result in the loss of nuanced details, reducing the effectiveness of guidance in complex pentesting phases.
\end{itemize}

The step-chaining process is used in a single pentest loop, consisting of the following three steps:

\begin{itemize}
\item \textbf{Command Generation}: Generating specific commands or actions for the pentest, tailored to the current context and objectives.
\item \textbf{Summarization of Current State}: Concisely summarizing the state of the pentest after each command, including outcomes and any changes in the scenario.
\item \textbf{Updation of a To-Do List}: Reflecting these changes in the to-do list, ensuring it remains current and relevant.
\end{itemize}

By chaining these steps together, a continuous flow of relevant information was maintained without exceeding the token limit. This method enabled more judicious token allocation, ensuring that each component was provided with sufficient space to remain effective without excessive verbosity.

\begin{figure}[H]
\centering
\includegraphics[width=0.9\columnwidth]{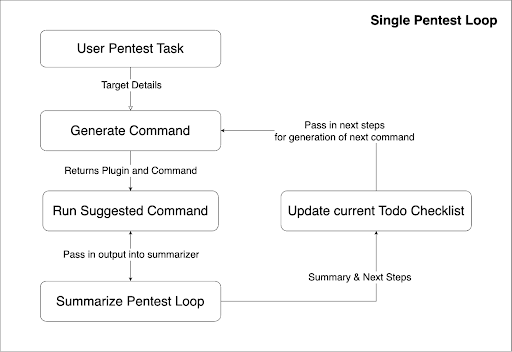}
\caption{Overview of a Single Pentest Loop covering the task from the user, generation of commands, running suggested commands followed by summarizing the pentest loop and updating the todo-list}
\label{fig2}
\end{figure}

The step chaining process (Fig.~\ref{fig2}) also ensured that the model kept a coherent and relevant context throughout the pentesting session, crucial for maintaining the quality and applicability of its guidance.

\subsection{Retrieval Augmented Generation}
Retrieval-Augmented Generation (RAG) plays a pivotal role in enhancing LLM Augmented Pentesting, particularly in addressing the challenge of providing up-to-date and accurate information on cybersecurity tools and practices \cite{lewis2021retrievalaugmentedgenerationknowledgeintensivenlp}.

One significant challenge in developing the tool was ensuring that the LLM had access to the latest information on the usage, syntax, and modules of complex cybersecurity tools and frameworks like Metasploit \cite{b9}. The dynamic nature of these tools, with frequent updates and new modules, means that any static knowledge base quickly becomes outdated.

To prepare the vector database for RAG in Pentest Copilot, a comprehensive dataset of the latest modules, scripts, and usage guides specifically from Metasploit and MSFVenom was collected. The dataset included structured information such as command syntax, use cases, and detailed documentation for these tools. This information was then processed to convert the textual data into vector representations, or embeddings, that capture the semantic meaning of the content. These embeddings were indexed and stored in the vector database, allowing for efficient similarity searches and quick retrieval of the most relevant and up-to-date information during LLM-driven pentesting sessions.

The RAG model serves the following purpose:

\begin{itemize}
\item \textbf{Query Processing}: When a user queries about a specific module or syntax, the LLM processes this query to understand the context and the specific information required.
\item \textbf{Information Retrieval}: The LLM then uses RAG to reach out to an external knowledge base, seeking the most relevant and current information related to the query.
\item \textbf{Data Integration}: The retrieved information is integrated with the LLM's internal knowledge to generate a comprehensive, formatted, accurate response.
\end{itemize}

\subsection{Preferred Tooling and Context}
In the LLM Augmented Pentesting approach, a researcher may prefer to use tools and techniques they are familiar with. To accommodate this, users can select their preferred tools, and these choices are then integrated into the system prompt. This customization allows the model to prioritize the user's chosen tools when suggesting commands. Additionally, the system prompt is designed to recommend alternative tools when deemed necessary. In such cases, the response provides a clear explanation for these recommendations, ensuring the user understands why an alternative tool might be more effective in a specific situation.

\medskip

\noindent\textbf{Example of Tool Preferences:}
\begin{itemize}
  \item \textbf{Reconnaissance:} \texttt{nmap} (TCP/UDP port scans), \texttt{masscan} (high-speed port sweeps), \texttt{naabu} (lightweight TCP/UDP enumeration)
  \item \textbf{Directory Enumeration:} \texttt{gobuster} (wordlist-based discovery), \texttt{dirb} (recursive directory brute-forcing)
  \item \textbf{Vulnerability Scanning:} \texttt{nikto} (web server scanning), \texttt{sqlmap} (automated SQL injection)
  \item \textbf{Service Enumeration:} \texttt{enum4linux} (SMB/LDAP enumeration), \texttt{snmpwalk} (SNMP data gathering)
\end{itemize}

\noindent When a user specifies, for example, \texttt{nmap} and \texttt{gobuster} as their preferred tools, the system prompt will have details related to the specified tools embedded (e.g.\ \texttt{-sC -sV -oA} for \texttt{nmap}, or \texttt{-u http://target -w /path/wordlist.txt} for \texttt{gobuster}) so that the AI’s suggested commands align with the user’s workflow.

\subsection{File Analysis with LLMs}
In pentesting, researchers frequently encounter files that, when analyzed, can lead to new discoveries. However, LLMs cannot directly understand the content of these files. Therefore, the information within these files must first be converted into a plaintext format that LLMs can interpret. 

The file analysis tool first checks the file format using the Linux ``file'' command. It then runs the appropriate analysis procedure based on the output of the file command. The tool also scans media files for any hidden file signatures, and recursively extracts and analyzes any additional files found.

The tool is currently designed to deal with such files by including large lines of text from the file in the output, limited by the allowed number of tokens to OpenAI's API \cite{b7,b8}.

For different file formats, the following contextual data is extracted:

\begin{itemize}
\item \textbf{ELF}: 
\begin{itemize}
\item Security Settings of the file: 
\begin{itemize}
\item NX bit
\item RELRO
\item Stack Canary
\item PIE
\end{itemize}
\item stdlib functions used in the file.
\item Symbols present in the binary.
\item Link Tye: Dynamic/Static.
\item Whether the binary is a shared object or not.
\item File Architecture.
\end{itemize}

\item \textbf{PE32/PE32+}:
\begin{itemize}
\item Dependencies of the executables.
\item Number of sections in the executable.
\item Security Settings present in the executable:
\begin{itemize}
\item Non-Executable Bit.
\item Stack Canary.
\item Structured Exception Handling.
\end{itemize}
\item Name of Libraries used by the executable.
\item Imported/Exported functions.
\end{itemize}

\item \textbf{XML/JSON/YAML}:
\begin{itemize}
\item For the configuration files, a minimalistic tree of all the parents and their possible child elements is created and structured to represent the hierarchical relationship between them.
\end{itemize}

\item \textbf{Media Files}:
\begin{itemize}
\item Exif data.
\item Any hidden signatures of other file types.
\end{itemize}

\item \textbf{Plain Text Files}:
\begin{itemize}
\item A python3 library ``Guesslang'' is used to detect any programming language. Depending on the detected language, function names are then extracted using regex \cite{b16}.
\item If no language is accurately detected, only very large lines (more than 100 words) are extracted and sent with the context.
\end{itemize}
\end{itemize}

The list of files that the File Analysis tool can analyze is currently limited to the following list:

\begin{itemize}
\item Binary files - ELF and PE32+ file formats for Linux and Windows
\item Media Files - MP3,MP4, PDF, PNG, JPEG etc. 
\item Configuration Files - XML, OpenVPN, JSON, and YAML.
\end{itemize}

\section{Infrastructure}

\subsection{Setting up Infrastructure}

\begin{figure}[H]
\centering
\includegraphics[width=0.9\columnwidth]{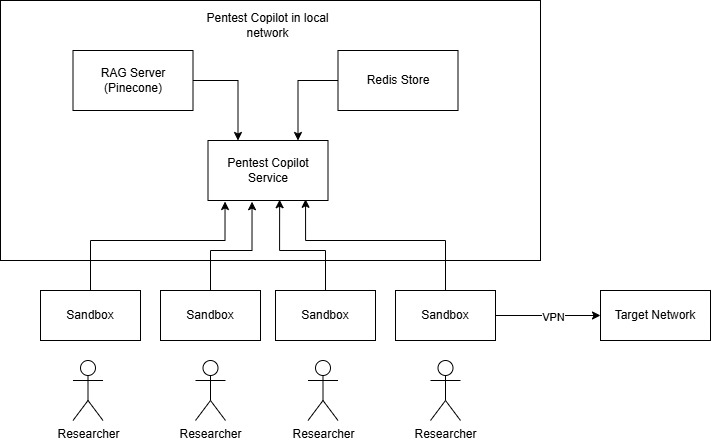}
\caption{Complete infrastructure of Pentest Copilot - All services and sandboxes are deployed within a single subnet, enabling seamless communication between containers, particularly when multiple researchers within the same organization may rely on the same server.}
\label{fig4}
\end{figure}

Pentest Copilot is designed to provide researchers with a robust platform for conducting seamless penetration testing sessions. To ensure uninterrupted testing, each session is allocated a dedicated sandbox compute instance, which can be the local system itself. This sandbox is pre-configured with a comprehensive suite of security tools (Table~\ref{tab:opensource_tools}), enabling users to perform a wide range of penetration testing activities without any dependency on external setups.

A critical aspect of effective penetration testing is access to graphical user interface (GUI) tools like Burp Suite. These tools play a vital role in dynamic analysis, making a full desktop experience essential. To accommodate this, Pentest Copilot integrates a complete desktop environment accessible directly through a browser session using the Virtual Network Computing (VNC) protocol. This setup allows users to seamlessly interact with GUI-based tools alongside command-line utilities, all within the same session.

Communication between the user's interface and the sandbox is managed through a secure terminal connection, using a socket-based model with Secure Shell (SSH) for command execution. This setup provides an interactive experience where users can run commands, receive real-time outputs, and maintain control over the pentesting process. Whether performing automated scans, executing manual exploits, or analyzing outputs, users experience consistent performance without connectivity disruptions.

\subsection{Choosing the right security tools}
The decision to integrate multiple open-source tools into Pentest Copilot was based on their open source nature and widespread use. As shown in Table~\ref{tab:opensource_tools}, several open-source tools were selected for integration in Pentest Copilot, ensuring a wide range of security testing capabilities.

For example, Ffuf is being used for directory brute forcing which provides the flexibility to choose the payload position and auto calibration for determining the baseline error response. These tools are frequently utilized by bug bounty hunters and security researchers to perform tasks essential to their work.\cite{b25} 

\begin{table*}[htpb]
\centering
\caption{List of OPEN SOURCE tools used by default in pentest copilot }
\label{tab:opensource_tools}
\begin{tabular}{|c|c|c|c|}
\hline
\textbf{Tool Name} & \textbf{GitHub Forks} & \textbf{GitHub Stars} & \textbf{Use Case} \\
\hline
Ffuf & 1.2k & 10.8k & Directory Brute Forcing \\
Feroxbuster* & 422 & 5k & \\
Dirsearch & 2.3k & 10.8k & \\
Wfuzz & 1.3k & 5.5k & \\
Gobuster & 1.2k & 8.6k & \\
\hline
Subfinder* & 1.2k & 8.8k & Subdomain Discovery Passive \\
Amass & 1.8k & 10.7k & \\
Findomain & 367 & 3.1k & \\
Assetfinder & 467 & 2.7k & \\
\hline
Shuffledns & 178 & 1.2k & \\
Puredns & 146 & 1.5k & Subdomain Brute Forcing \\
dnsx & 235 & 1.8k & \\
\hline
THC-Hydra* & 1.9k & 8.7k & Credentials Brute Forcing \\
Patator & 788 & 3.4k & \\
Crowbar & 328 & 1.3k & \\
\hline
Wpscan & 1.3k & 8.1k & CMS Scanners \\
Drupwn & 134 & 549 & \\
CMSmap & 249 & 939 & \\
\hline
Nmap* & 2.3k & 8.8k & Port Scanning \\
Naabu & 482 & 4k & \\
Smap & 234 & 2.7k & \\
Masscan & 3.3k & 22.2k & \\
\hline
WayBackURLs & 434 & 3k & Spider/Crawling \\
Gau & 401 & 3.3k & \\
xnlLinkFinder & 130 & 963 & \\
Waymore & 146 & 1.2k & \\
Katana & 434 & 3k & \\
Gospider & 337 & 2.3k & \\
\hline
Whatsweb & 906 & 5k & Fingerprinting Technologies \\
Httpx & 756 & 6.4k & \\
\hline
Sqlmap* & 5.6k & 29.6k & Vulnerability Exploitation \\
Ghauri & 201 & 1.7k & \\
GraphQLmap & 188 & 1.2k & \\
Dalfox & 379 & 3.1k & \\
SecretFinder & 331 & 1.6k & \\
\hline
\end{tabular}
\end{table*}

\subsection{VPN Configuration}
Pentest Copilot is also made capable of targeting entities located on private subnets which is a common use-case in pentesting engagements.

This is achieved through establishing an OpenVPN server. This ensures secure and controlled access to the private network.

Users may also establish a secure connection to remote networks using a tool called Chisel in case they don't have a VPN server setup. This ensures flexible and secure communication between Pentest Copilot and the target environment.

\subsection{Mitigating Possible Malicious Use}
Effective security policies are essential to ensure that Pentest Copilot remains a safe and ethical tool for penetration testing, regardless of the deployment environment. Whether deployed on cloud infrastructure, an on-premises network, or within an organization's internal setup, the platform must be safeguarded against potential misuse or exploitation. 

The implemented security measures, as outlined in Table~\ref{tab4}, ensure that Pentest Copilot can be securely deployed in various environments while maintaining ethical testing standards.

\begin{table}[H]
\centering
\caption{Risk and Mitigation Analysis}
\begin{tabular}{|p{1.9cm}|p{2.4cm}|p{2.4cm}|}
\hline
\textbf{Type} & \textbf{Risk} & \textbf{Mitigation} \\
\hline
Malicious use of compute resources & Password cracking or Crypto Mining on sandbox infrastructure & CPU Usage Monitoring and Execution caps on resources. \\
\hline
Malicious use of Pentest Copilot & Users targeting systems which one is not authorized to test on from sandbox instance & By default, implement no egress traffic to public internet \\
\hline
Exploitation of Pentest Copilot & Leaked API keys & Secrets Management and Key Rotation \\
\hline
Exploitation of Pentest Copilot & RCE on Pentest Copilot instance & Rigorous and Regular Pentesting and Remediation \\
\hline
\end{tabular}
\label{tab4}
\end{table}

\section{Result Analysis and Justification}

\subsection{Performance}
The journey of Pentest Copilot's development has seen remarkable improvements in performance, particularly in the areas of response time, accuracy, and functionality. These enhancements have significantly elevated the tool's utility for pentesters, streamlining their workflow and providing more precise assistance.
The summarized flow of Pentest Copilot is detailed in Algorithm~\ref{alg:llm_pentest_flow}.

\setlength{\algomargin}{2em}

\begin{algorithm}[htbp]
\caption{LLM-Augmented Pentesting Flow (Pentest Copilot)}
\label{alg:llm_pentest_flow}
\SetAlgoLined

\KwIn{User input query $Q$, Target system $T$, Tool preferences \textit{Prefs}}
\KwOut{Context-aware pentesting recommendations and commands}

\textbf{Initialize:} Load GPT-4-Turbo with System Prompt for Penetration Testing\;

\uIf{$Q$ includes prior context}{
    Inject prior context into prompt\;
}
\Else{
    Trigger preliminary recon phase (domain/IP scan)\;
}

\textbf{Configure plugins:} WebSearch, RunBash, NetcatListener, GeneratePayload, GenericResponse\;
Embed tool preferences \textit{Prefs} into system prompt\;

\textbf{Retrieval-Augmented Generation (RAG):}\;
Query vector DB for latest tool syntax and usage modules\;
Merge retrieved info with LLM context\;

\While{Pentest session is active}{
    Generate command via LLM based on current context\;
    Execute command on sandbox via SSH\;
    Summarize output and update To-Do list\;
    Re-inject summary into next prompt\;
}

\uIf{file is uploaded}{
    Identify file type, extract text, and convert to plaintext\;
    Incorporate extracted insights into prompt context\;
}

\textbf{Result Delivery:}\;
Display command outputs, insights, and To-Do updates in client UI\;

\textbf{Session Cleanup:}\;
Store artifacts and final command logs\;
Terminate session with security checks and cleanup\;

\Return Final findings summary, command logs, and updated To-Do checklist\;

\end{algorithm}

Without any optimizations, Pentest Copilot took an average of 8-10 minutes to perform a series of critical tasks:

\begin{itemize}
\item Generate a Command: Crafting a specific command relevant to the current stage of the pentest.
\item Summarize the Current Pentest State: Providing a concise overview of the pentest's current status.
\item Update the To-Do List: Reflecting new tasks or findings in the pentest checklist.
\end{itemize}

This duration was a bottleneck, especially in scenarios where quick decision-making and rapid response were crucial.

Transitioning to GPT-4-Turbo, with its 128,000-token capacity, reduced this time to 4--5 minutes, a nearly 50\% improvement, while achieving 60\% command accuracy compared to GPT-4's 50\% and GPT-3.5-Turbo's 40\%. This enhancement, validated through a testbenching framework simulating real-world pentesting scenarios like remote code execution on a ``boot2root box,'' demonstrates the superior suitability of advanced GPT models for dynamic, context-rich pentesting workflows.

To ensure the tool's outputs remain current and reliable, Retrieval-Augmented Generation (RAG) was integrated, drawing from a curated vector database of the latest Metasploit and MSFVenom modules. This approach mitigates hallucinations by supplementing the LLM's internal knowledge with up-to-date tool syntax and usage guides, resulting in more precise and relevant command generation. Complementing this, Pentest Copilot's file analysis capability transforms non-textual inputs---such as ELF and PE32 binaries, configuration files, and media---into plaintext, extracting actionable data like security settings, dependencies, and Exif metadata. This functionality enables the LLM to reason over complex pentesting artifacts, streamlining vulnerability identification and post-pentest analysis, thus proving its adaptability to diverse operational inputs.

The infrastructure supporting Pentest Copilot further justifies its effectiveness, by seamlessly integrating with an ``exploit-box'' server which has pre-installed tools like Nmap and Sqlmap, alongside VNC for GUI tools like Burp Suite. Socket-based communication via SSH ensures session continuity and real-time interactivity, while security measures---CPU monitoring, restricted internet access, and VPN support (OpenVPN and Chisel)---safeguard against misuse and enable secure targeting of private subnets.

A significant addition to Pentest Copilot is the introduction of file analysis functionality. This feature enables pentesters to efficiently analyze findings and data collected during a pentest. It simplifies the process of sifting through complex datasets, logs, or reports, allowing for quicker identification of vulnerabilities, anomalies, or patterns. This capability is particularly valuable in post-pentest reviews, where thorough analysis is key to understanding the effectiveness of the pentest and planning future security strategies.

A comparative analysis was conducted across five leading AI-driven pentesting tools to evaluate their feature sets and operational capabilities. Pentest Copilot was benchmarked against PentestGPT, GPTPEN, BreachStorm, and AUTOPOWNED, examining aspects such as environment type, test orchestration, prompt customizability, file and credential handling, and goal‐oriented reasoning. This analysis highlights Pentest Copilot’s strengths—full desktop environments, multi-step orchestration, extensive plugin support, and a reinforcement-driven reasoning loop—while contrasting each tool’s limitations and focus areas (Table \ref{tab:compare_all}).

\begin{table*}[htbp]
  \centering
  \caption{Comparative Analysis of AI-Driven Pentesting Tools}
  \begin{tabular}{|p{2cm}|p{2.5cm}|p{2.5cm}|p{2.2cm}|p{2.2cm}|p{2.6cm}|}
    \hline
    \textbf{Feature / Tool} & \textbf{Pentest Copilot} & \textbf{PentestGPT} & \textbf{GPTPEN} & \textbf{BreachStorm} & \textbf{Autopowned} \\
    \hline
    Environment Type 
      & Full desktop environment via ephemeral VNC 
      & VSCode + Terminal-Notebook 
      & SSH shell, editor-enabled private 
      & Localized cloud Linux agents 
      & Pre-configured lab pipelines \\
    \hline
    Test Orchestration 
      & Manual and multi-step triggered 
      & Semi-automated 
      & Manual only 
      & Not exposed to users 
      & Not exposed to users \\ 
    \hline
    Prompt Customizability 
      & Multi-agent chain (explorer + investigator support) 
      & Not orchestrated 
      & Not orchestrated 
      & Static notebooks only 
      & Static notebooks only \\ 
    \hline
    File \& Credential Handling 
      & GUI, API, Nmap, Metasploit, Certipy, custom scripts 
      & Not supported 
      & Not supported 
      & Static analysis only 
      & Static analysis only \\ 
    \hline
    Goal / Intent Reasoning 
      & LLM with reinforcement feedback loop 
      & No goal-based reasoning support 
      & Basic LLM request-only 
      & No permanent sessions 
      & No goal reasoning \\
    \hline
  \end{tabular}
  \label{tab:compare_all}
\end{table*}

\subsection{Shortcomings}
Pentest Copilot's current capabilities are limited to the underlying LLM. This constraint implies that the system may not fully comprehend or execute complex operations associated with sophisticated testing tools or methodologies, potentially hindering its performance in scenarios demanding advanced capabilities beyond the scope of the LLM. These limitations include, but are not limited to:

\begin{itemize}
\item Not being able to mount zero-day attacks
\item Unable to bypass security controls which are not known to the LLM
\item Using attack patterns which may have been overused a lot resulting in easy detection by security systems.
\end{itemize}

At present, the current approach lacks in-depth awareness of the intricate operations of closed-source and paid frameworks/tools. For example, the tool is currently unable to properly instruct a new Cobalt Strike user on how to set up Beacons, Proxies and mount lateral traversal attacks and correlate each Beacon's position and importance in the network. The platform's capabilities may be constrained in dealing with the specific functionalities and nuances of these tools, potentially limiting its effectiveness in certain advanced testing scenarios which touch upon red teaming for example, malware writing, bypassing EDR (Endpoint Detection and Response) or advanced Active Directory attack scenarios.

\section{Conclusion}
Pentest Copilot is set to advance by integrating with diverse knowledge bases for technology-specific expert advice. Future versions will feature fine-tuned models for pentesting and red teaming activities, enhancing capabilities in exploit writing and understanding closed-source industry tools in the pentesting and red-teaming spaces. The LLM Augmented Pentesting approach will involve maintaining a dynamic buffer of identified misconfigurations during targeting, allowing real-time adaptation and pattern recognition.

Additionally, a red team-specific knowledge base will be developed, covering both non-technical security issues (e.g., phishing) and technology-heavy misconfigurations (e.g., Group Policy Objects misconfigurations in Active Directory). These strategic enhancements aim to make Pentest Copilot an agile, expert, and indispensable tool for ethical hackers and penetration testers.

\section*{Acknowledgment}
The authors would like to thank Siddharth Johri for his work on VPN configuration and deployment, and Devang Solanki and Tuhin Bose for their contributions to security and functionality testing, including identifying errors in the solution. We also extend our gratitude to Bhavarth Karmarkar for his efforts in developing the file analysis solution.

The authors would like to express their gratitude to the developers and maintainers of the various open-source tools that were made available to Pentest Copilot. Although we did not use these tools directly in our research, they played a crucial role in enabling the framework to access a wide range of capabilities for comprehensive security testing and analysis:

\begin{itemize}
\item \textbf{Directory Brute Forcing}: ffuf, Feroxbuster, Dirsearch, Wfuzz, and Gobuster
\item \textbf{Subdomain Discovery (Passive)}: Subfinder, Amass, Findomain, and Assetfinder
\item \textbf{Subdomain Brute Forcing}: Shuffledns, Puredns and dnsx
\item \textbf{Credentials Brute Forcing}: THC-Hydra, Patator and Crowbar
\item \textbf{CMS Scanners}: Wpscan, Drupwn and CMSmap
\item \textbf{Port Scanning}: Naabu, Smap, Masscan, and Nmap
\item \textbf{Spidering/Crawling}: Gau, xnLinkFinder, Waymore, Katana, Gospider, and waybackurls
\item \textbf{Fingerprinting Technologies}: Httpx and WhatWeb
\item \textbf{Vulnerability Exploitation}: Sqlmap, Ghauri, GraphQLMap, Dalfox, SecretFinder
\end{itemize}

\bibliographystyle{unsrt}
\bibliography{references}

\begin{thebibliography}{10}

\bibitem{tann2023usinglargelanguagemodels}
Wesley Tann, Yuancheng Liu, Jun~Heng Sim, Choon~Meng Seah, and Ee-Chien Chang.
\newblock Using large language models for cybersecurity capture-the-flag challenges and certification questions, 2023.

\bibitem{b18}
Cybersecurity: Market data and analysis.
\newblock Handelsblatt Live, 2023.

\bibitem{Happe_2023}
Andreas Happe and Jürgen Cito.
\newblock Getting pwn’d by ai: Penetration testing with large language models.
\newblock In {\em Proceedings of the 31st ACM Joint European Software Engineering Conference and Symposium on the Foundations of Software Engineering}, ESEC/FSE ’23, page 2082–2086. ACM, November 2023.

\bibitem{peng2023impactaideveloperproductivity}
Sida Peng, Eirini Kalliamvakou, Peter Cihon, and Mert Demirer.
\newblock The impact of ai on developer productivity: Evidence from github copilot, 2023.

\bibitem{Pratama_2024}
Derry Pratama, Naufal Suryanto, Andro~Aprila Adiputra, Thi-Thu-Huong Le, Ahmada~Yusril Kadiptya, Muhammad Iqbal, and Howon Kim.
\newblock Cipher: Cybersecurity intelligent penetration-testing helper for ethical researcher.
\newblock {\em Sensors}, 24(21):6878, October 2024.

\bibitem{zhang2023doesllmgeneratesecurity}
Ying Zhang, Wenjia Song, Zhengjie Ji, Danfeng, Yao, and Na~Meng.
\newblock How well does llm generate security tests?, 2023.

\bibitem{deng2024pentestgptllmempoweredautomaticpenetration}
Gelei Deng, Yi~Liu, Víctor Mayoral-Vilches, Peng Liu, Yuekang Li, Yuan Xu, Tianwei Zhang, Yang Liu, Martin Pinzger, and Stefan Rass.
\newblock Pentestgpt: An llm-empowered automatic penetration testing tool, 2024.

\bibitem{alshehri2024breachseekmultiagentautomatedpenetration}
Ibrahim Alshehri, Adnan Alshehri, Abdulrahman Almalki, Majed Bamardouf, and Alaqsa Akbar.
\newblock Breachseek: A multi-agent automated penetration tester, 2024.

\bibitem{xu2024autoattackerlargelanguagemodel}
Jiacen Xu, Jack~W. Stokes, Geoff McDonald, Xuesong Bai, David Marshall, Siyue Wang, Adith Swaminathan, and Zhou Li.
\newblock Autoattacker: A large language model guided system to implement automatic cyber-attacks, 2024.

\bibitem{pearce2022examiningzeroshotvulnerabilityrepair}
Hammond Pearce, Benjamin Tan, Baleegh Ahmad, Ramesh Karri, and Brendan Dolan-Gavitt.
\newblock Examining zero-shot vulnerability repair with large language models, 2022.

\bibitem{schwartz2019autonomouspenetrationtestingusing}
Jonathon Schwartz and Hanna Kurniawati.
\newblock Autonomous penetration testing using reinforcement learning, 2019.

\bibitem{b7}
OpenAI.
\newblock Models - openai api.
\newblock \url{https://platform.openai.com/docs/models/}.
\newblock Accessed: 2023-02-02.

\bibitem{b8}
OpenAI.
\newblock Gpt-4.
\newblock \url{https://openai.com/research/gpt-4}.
\newblock Accessed: 2023-06-30.

\bibitem{b11}
Hack~The Box.
\newblock Hackthebox: Hacking training for the best.
\newblock \url{https://www.hackthebox.com/}.

\bibitem{b12}
TryHackMe.
\newblock Tryhackme: A free online platform for learning cyber security.
\newblock \url{https://tryhackme.com/}.

\bibitem{b13}
VulnHub.
\newblock Vulnhub: Vulnerable machines for practicing hacking.
\newblock \url{https://www.vulnhub.com/}.

\bibitem{liu2024jailbreakingchatgptpromptengineering}
Yi~Liu, Gelei Deng, Zhengzi Xu, Yuekang Li, Yaowen Zheng, Ying Zhang, Lida Zhao, Tianwei Zhang, Kailong Wang, and Yang Liu.
\newblock Jailbreaking chatgpt via prompt engineering: An empirical study, 2024.

\bibitem{b9}
Rapid7.
\newblock Metasploit framework.
\newblock \url{https://www.metasploit.com/}.
\newblock Accessed: 2023-07-30.

\bibitem{wei2023chainofthoughtpromptingelicitsreasoning}
Jason Wei, Xuezhi Wang, Dale Schuurmans, Maarten Bosma, Brian Ichter, Fei Xia, Ed~Chi, Quoc Le, and Denny Zhou.
\newblock Chain-of-thought prompting elicits reasoning in large language models, 2023.

\bibitem{lewis2021retrievalaugmentedgenerationknowledgeintensivenlp}
Patrick Lewis, Ethan Perez, Aleksandra Piktus, Fabio Petroni, Vladimir Karpukhin, Naman Goyal, Heinrich Küttler, Mike Lewis, Wen tau Yih, Tim Rocktäschel, Sebastian Riedel, and Douwe Kiela.
\newblock Retrieval-augmented generation for knowledge-intensive nlp tasks, 2021.

\bibitem{b16}
Yoeo.
\newblock Guesslang: Detect the programming language of a source code.
\newblock GitHub, 2024.

\bibitem{b25}
J.~K. Binu.
\newblock Best directory brute-forcing tools for beginner bug hunters.
\newblock Medium, 2024.

\end{thebibliography}

\end{document}